\documentclass[aip,jmp,nofootinbib,preprint]{revtex4-1}

\usepackage{amsmath}
\usepackage{amsthm}
\usepackage{amssymb}
\usepackage{amscd}
\usepackage{bm}
\usepackage{latexsym}
\usepackage{mathtools}
\usepackage{xcolor}

\usepackage{graphicx}
\usepackage{epsfig}
\usepackage{epstopdf}
\usepackage[caption=false]{subfig}

\renewcommand{\vec}[1]{\bar{#1}}

\providecommand{\eqref}[1]{(\ref{#1})}
\providecommand{\vanish}[1]{}

\begin{document}

\preprint{ }

\title{CONTINUOUS AND OPTIMALLY COMPLETE DESCRIPTION OF CHEMICAL ENVIRONMENTS USING SPHERICAL BESSEL DESCRIPTORS}

\author{Emir Kocer}
\affiliation{Department of Mechanical Engineering, Bogazici University, Istanbul, TURKEY}
\author{Jeremy K. Mason}
\affiliation{Department of Materials Science and Engineering, University of California Davis, CA, USA}
\author{Hakan Erturk}
\affiliation{Department of Mechanical Engineering, Bogazici University, Istanbul, TURKEY}

\begin{abstract}
Recently, machine learning potentials have been advanced as candidates to combine the high-accuracy of electronic structure methods with the speed of classical interatomic potentials. A crucial component of a machine learning potential is the description of local atomic environments by some set of descriptors. These should ideally be invariant to the symmetries of the physical system, twice-differentiable with respect to atomic positions (including when an atom leaves the environment), and complete to allow the atomic environment to be reconstructed up to symmetry. The stronger condition of optimal completeness requires that the condition for completeness be satisfied with the minimum possible number of descriptors. Evidence is provided that an updated version of the recently proposed Spherical Bessel (SB) descriptors satisfies the first two properties and a necessary condition for optimal completeness. The Smooth Overlap of Atomic Position (SOAP) descriptors and the Zernike descriptors are natural counterparts of the SB descriptors and are included for comparison. The standard construction of the SOAP descriptors is shown to not satisfy the condition for optimal completeness, and moreover is found to be an order of magnitude slower to compute than the SB descriptors.

\end{abstract}

\maketitle

\section{Introduction}
\label{sec:introduction}

Machine learning potentials (MLPs) have recently become a subject of interest in computational materials science \cite{behler2016perspective}, potentially offering the accuracy of electronic structure techniques like density functional theory (DFT) without the associated computational cost. An MLP effectively learns to reproduce the potential energy surface (PES), i.e., the hypersurface that defines the potential energy of an atomic system as a function of the atomic positions. While the reliability of atomistic simulations including molecular dynamics (MD) depends on the accuracy of the PES, their usefulness to study complex phenomena is limited by the accessible time and length scales; in practice this makes the computational cost of an MD simulation nearly as much a concern as the accuracy. Recent studies \cite{artrith2012high,bartok2018machine,botu2016machine} suggest that MLPs can achieve a favorable combination of performance and accuracy that is provided by neither classical force fields nor electronic structure calculations.

Machine learning (ML) algorithms that have been employed to construct MLPs include artificial neural networks (ANNs) \cite{blank1995neural,behler2007generalized}, support vector machines (SVMs) \cite{balabin2011support} and Gaussian processes (GPs) \cite{bartok2010gaussian}. Regardless of the algorithm, MLPs rely on the reasonable assumption that the energy of an atom is a multidimensional function of the relative positions of the neighboring atoms. This atom-centered approach \cite{behler2007generalized} enables the total energy $E$ of a system to be calculated by summing over all individual atomic energies $E_i$ as
\begin{equation*} 
E=\sum_{i} E_i
\end{equation*}
and reduces the problem to one involving a local atomic environment. This environment is usually encoded as a set of scalars, known as \textit{descriptors}, that serve as the inputs for the atom-centered MLPs. Faber et al.\ \cite{faber2017prediction} carried out a systematic study of how the choice of descriptors and ML algorithm can affect the accuracy of an MLP by testing a variety of combinations. They found that the choice of descriptors could affect the accuracy more than the regression scheme, justifying the effort spent over the last decade in developing the many competing descriptors available in the literature \cite{bartok2010gaussian,behler2007generalized,bartok2013representing,novotni20033d,rupp2012fast,glielmo2017accurate,huo2017unified}. Of these, the Behler-Parinello (BP) symmetry functions \cite{behler2007generalized} and the Smooth Overlap of Atomic Position (SOAP) descriptors \cite{bartok2013representing} are some of the most frequently used, and have been employed in MLPs that achieve the accuracy of electronic structure methods in a variety of applications \cite{rowe2018development,deringer2017machine,artrith2016implementation,liu2018constructing}. Afterwards, Khorshidi et al.\ proposed to use the Zernike polynomials\cite{novotni20033d,khorshidi2016amp} and the neighbor density function of Bartok et al.\ \cite{bartok2010gaussian} to construct the Zernike descriptors, and reported comparable results. Recently, Kocer et al.\ \cite{kocer2019novel} proposed to use the spherical Bessel functions with a closely related procedure to construct the Spherical Bessel (SB) descriptors. These were found to allow construction of MLPs significantly more accurate than those using the BP symmetry functions, and of comparable accuracy to but an order of magnitude faster to evaluate than those using the SOAP descriptors.

Any set of descriptors should satisfy a number of mathematical properties to not constrain the ability of the ML algorithm to approximate the PES. First, it is desirable from a computational standpoint that they be invariant to the symmetries of the physical system (i.e., translations, rotations, inversions and permutation of atomic labels) to reduce the domain of the PES and the number of training examples required. More subtle but perhaps more important is that the descriptors be similar but distinct for similar but distinct atomic environments: If the descriptors are not similar, the MLP would not likely be continuous, and if the descriptors are not distinct, the MLP would not be able to reproduce potentially significant features of some physical systems. This is closely related to the concept of \emph{completeness}, here defined as the condition that the space of all local atomic environments that are not related by physical symmetries is smoothly embedded into the space of descriptors. This is desireable because, e.g., a set of descriptors that is complete allows the atomic environment to be reconstructed up to symmetry. The stronger condition of \emph{optimal completeness} requires that the embedding into the space of descriptors always be achieved with the minimum number of descriptors, and is highly desireable for computational reasons. Finally, the descriptors should be twice-differentiable to allow for continuity of forces and elastic constants, contain few adjustable parameters to help with transferrability of the potentials, and be numerically efficient to evaluate. To the extent of our knowledge, none of the descriptors available in the literature fulfills all of these requirements.

This paper presents an updated version of the SB descriptors (requiring only a change in indexing) that makes them continuous with respect to atomic displacements. A necessary condition for optimal completeness is then formulated using the Rank Theorem \cite{krantz2012implicit}. The SB descriptors are found to satisfy this condition, whereas the power spectrum coefficients used in the construction of the SOAP descriptors do not. Finally, the accuracy and efficiency of the SB descriptors in a proof-of-concept MLP are compared to several of the alternatives available in the literature.

\section{Spherical Bessel descriptors}
\label{sec:SB_desc}

Following a similar procedure to our recent study \cite{kocer2019novel}, an atomic neighbor density function
\begin{equation}
\rho^k(\vec{r}) = \sum_{j} w^k_{ij} \delta(\vec{r}-\vec{r}_{ij})
\label{eq:density_function}
\end{equation}
is first defined for a central atom $i$, where $\vec{r}_{ij}$ are the relative position vectors of each neighbor $j$ with respect to $i$. The weight factor $w^k_{ij}$ could be used to specify the species of atoms $i$ and $j$ in a multi-component system, but is assumed to be one in this study. The neighbor density function $\rho(\vec{r})$ is projected onto a set of orthonormal basis functions on the ball of radius $r_c$, giving an expansion of the form
\begin{equation} \label{eq:expansion}
\rho(\vec{r}) \approx \sum_{n = 0}^{n_\mathrm{max}} \sum_{l = 0}^{n} \sum_{m = -l}^{l} c_{nlm} g_{n-l,l}(r) Y^m_l(\theta,\phi)
\end{equation}
where $g_{nl}(r)$ is a radial basis function, $Y^m_{l}(\theta,\phi)$ is a spherical harmonic, and $n_{max}$ specifies the order of the approximation. While many functions could be used for the $g_{nl}(r)$, the one for the SB descriptors begins with the linear combination
\begin{equation*} \label{fnl}
f_{nl}(r)=a_{nl}j_l\left (r \frac{u_{ln}}{r_c}\right )+b_{nl}j_l\left (r \frac{u_{l,n+1}}{r_c}\right )
\end{equation*}
where $a_{nl}$ and $b_{nl}$ are constants, $j_l(r)$ is the $l$th spherical Bessel function of the first kind, $u_{ln}$ is the $(n+1)$th nonzero root of $j_l(r)$, and $r_c$ is the cutoff radius. The condition $f_{nl}(r_c)=0$ is satisfied by definition, and $a_{nl}$ and $b_{nl}$ can surprisingly be chosen to simultaneously satisfy the conditions $f'_{nl}(r_c)=0$ and $f''_{nl}(r_c)=0$, i.e., to make the radial basis functions twice differentiable at the cutoff radius. Along with normalization, this leads to
\begin{equation*} 
f_{nl}(r)= \bigg( \frac{1}{r_c^3} \frac{2}{u_{ln}^2 + u_{l,n+1}^2} \bigg)^{1/2} \left [\frac{u_{l,n+1}}{j_{l+1}(u_{ln})}j_l\left (r \frac{u_{ln}}{r_c}\right ) - \frac{u_{ln}}{j_{l+1}(u_{l,n+1})}j_l\left (r \frac{u_{l,n+1}}{r_c} \right )\right ].
\end{equation*}
The radial basis functions $g_{nl}(r)$ are then obtained by applying a Gram-Schmidt process to the $f_{nl}(r)$ for $0 \leq n \leq n_{max}$. A detailed derivation of the $g_{nl}(r)$ and explicit recursion relations that allow them to be efficiently evaluated are provided in the supplementary material. Given the radial and angular basis functions in Eq.\ \ref{eq:expansion}, the expansion coefficients $c_{nlm}$ for the $i$th atom are calculated from the relative spherical coordinates $(r_{ij},\theta_{ij},\phi_{ij})$ of the neighboring atoms as
\begin{equation}
c_{nlm}=\sum_{j}g_{n-l,l}(r_{ij})Y^{m*}_{l}(\theta_{ij}, \phi_{ij})
\label{eq:cnlm}
\end{equation}
by means of the standard orthogonality relations. The power spectrum $p_{nl}$ obtained from
\begin{equation}
p_{nl}=\sum_{m=-l}^{l} c^{*}_{nlm} c_{nlm}
\label{eq:pnl}
\end{equation}
 then comprises an infinite set of real-valued numbers that are used as the local structural descriptors. They are invariant to translations by the use of relative spherical coordinates, and to permutations of atomic labels by the construction of the neighbor density function in Eq.\ \ref{eq:density_function}. Invariance to rotations and inversions can be seen by substituting Eq.\ \ref{eq:cnlm} into Eq.\ \ref{eq:pnl} and reordering the summations to find 
\begin{equation*} 
p_{nl} = \sum_{j} \sum_{k} g_{n-l,l}(r_{j}) g_{n-l,l}(r_{k}) \sum_{m=-l}^{l} \bigg[  Y^m_{l}(\theta_{j},\phi_{j}) Y^{m*}_{l}(\theta_{k},\phi_{k}) \bigg]
\end{equation*}
where the subscript $i$ is suppressed for clarity. The spherical harmonic addition theorem \cite{arfken1985mathematical} allows this to be reduced to
\begin{equation}
p_{nl} = \frac{2 l + 1}{4 \pi} \sum_{j} \sum_{k} g_{n-l,l}(r_{j}) g_{n-l,l}(r_{k}) P_l(\cos \gamma_{jk})
\label{eq:pnl_reduced}
\end{equation}
where $P_l$ is the Legendre polynomial of order $l$ and $\gamma_{jk}$ is the triplet angle between atoms $i$, $j$ and $k$. Since the radial distances and triplet angles that constitute the independent variables in Eq.\ \ref{eq:pnl_reduced} are invariant to rotations and inversions, the $p_{nl}$ necessarily have the same property. A second reason to consider Eq.\ \ref{eq:pnl_reduced} as defining the $p_{nl}$ is that Eq.\ \ref{eq:pnl_reduced} is much more efficient to evaluate than Eqs. \ref{eq:cnlm} and \ref{eq:pnl}.

\begin{figure}
\center
\subfloat[]{%
	\label{subfig:ours-old}{%
		\includegraphics[width=0.35\textwidth]{%
			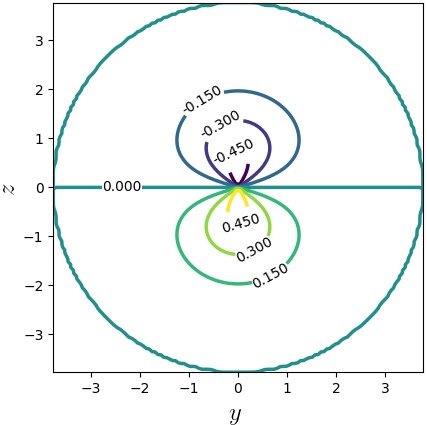}}} \quad
\subfloat[]{%
	\label{subfig:soap}{%
		\includegraphics[width=0.35\textwidth]{%
			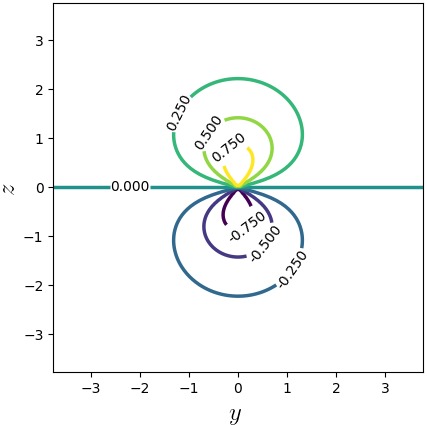}}}\quad
\subfloat[]{%
	\label{subfig:ours-new}{%
		\includegraphics[width=0.35\textwidth]{%
			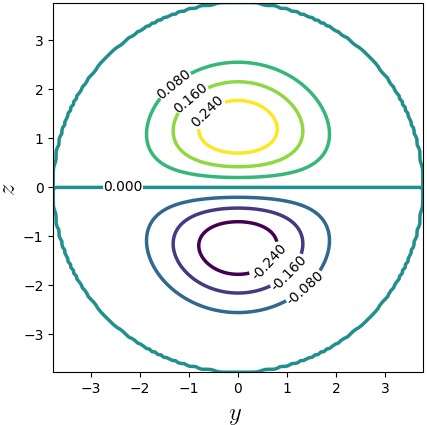}}} \quad
\subfloat[]{%
	\label{subfig:zernike}{%
		\includegraphics[width=0.355\textwidth]{%
			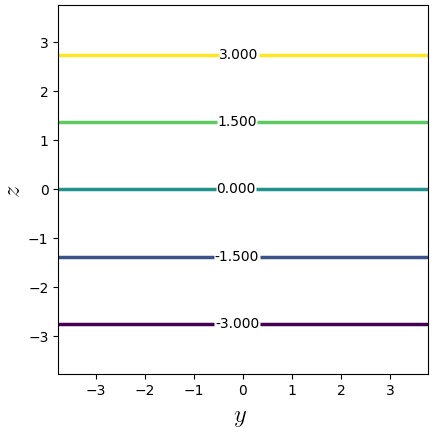}}}
\caption{\label{fig:basis_functions}Contour plots on the $yz$ plane of the basis functions used to construct (a) the previous SB descriptors \cite{kocer2019novel} for $n=0$, $l=1$ and $m=0$ ($g_{00} Y_{10}$), (b) the SOAP descriptors \cite{szlachta2014accuracy} for $n=0$, $l=1$ and $m=0$, (c) the current SB descriptors for $n=1$, $l=1$ and $m=0$ ($g_{01} Y_{10}$), and (d) the Zernike descriptors \cite{khorshidi2016amp} for $n=1$, $l=1$ and $m=0$. There is a visible discontinuity at the origin in (a) and (b).}
\end{figure}

The original definition\cite{kocer2019novel} of the radial basis functions $g_{nl}(r)$ included the constraint $l = 0$, removing the coupling between the angular and radial parts in Eq.\ \ref{eq:expansion}. While this significantly simplified the evaluation without an observable effect on the accuracy of the MLP for that specific set of training data, this also introduced discontinuities around the origin (near the central atom) in the basis functions in Eq.\ \ref{eq:expansion} for odd $l$ (Fig.\ \ref{subfig:ours-old}). Precisely the same issue occurs for the basis functions used in the SOAP descriptors (Fig.\ \ref{subfig:soap}), and is likely the motivation for using a superposition of Gaussians in the neighbor density function \cite{bartok2013representing} to smooth over the discontinuity. This approach comes at the price of expensive numerical integrations when evaluating the descriptors though, and introduces additional adjustable parameters. The proposed SB descriptors instead use basis functions that are twice differentiable everywhere (Fig.\ \ref{subfig:ours-new}), allowing the neighbor density function to be written as a superposition of Dirac delta functions and the descriptors to be calculated at least an order of magnitude faster. Specifically, the MATLAB implementations of the SB descriptors and the SOAP descriptors provided in the supplementary material respectively required $1.15$ and $16.1$ seconds on a 2.60GHz CPU to calculate a comparable number of descriptors for $1000$ atomic environments. The implementation of the SOAP descriptors follows that of standard references \cite{bartok2013representing,szlachta2014accuracy}, and employed a custom implementation of the double exponential integration technique \cite{takahasi1974double} to accelerate the numerical integration.

Part of the appeal of the SOAP descriptors is that they leave a number of choices up to the practitioner. With specific regard to computational efficiency, a recent publication uses several approximations and a particular choice of radial basis functions to calculate the SOAP descriptors without numerical integration \cite{caro2019optimizing}. While this approach is indeed more efficient, the effect of the required approximations is unclear, all of the basis functions with odd values of $l$ contain discontinuities at the origin of the type in Fig.\ \ref{subfig:soap}, and the orthogonalization of the radial basis functions does not include the appropriate weight factor for the spherical coordinate system, propagating a mistake made in the prior literature \cite{bartok2013representing}. For these reasons, this particular variation of the SOAP descriptors will not be considered further.

The basis functions of the Zernike descriptors \cite{canterakis19993d,khorshidi2016amp} are known as the Zernike polynomials, are rotationally-invariant orthogonal polynomials in $x$, $y$ and $z$, and do not contain any discontinuity within the cutoff sphere (Fig.\ \ref{subfig:zernike}). While the Zernike polynomials have several other desireable properties, they do not vanish at the cutoff radius and actually oscillate most rapidly there for higher $n$ and $l$. This effectively concentrates their ability to resolve atomic positions in the regions furthest from the central atom, whereas intuition suggests that the dependence of the potential energy on atomic position should be weakest there. While using a cutoff function in the definition of the neighbor density function does make the descriptors differentiable as atoms leave the environment, this also discards much of the information from the boundary region where the Zernike polynomials are most sensitive and introduces additional adjustable parameters. The relative performance of the Zernike descriptors is considered further in Sec.\ \ref{sec:efficiency}.

\section{Completeness}

Of the desirable mathematical properties of a set of descriptors identified in Sec.\ \ref{sec:introduction}, the most difficult one to establish is completeness. This word means different things for the basis functions and the descriptors though. For the basis functions, completeness indicates that the expansion in Eq.\ \ref{eq:expansion} is over a complete orthonormal basis, i.e., that the expansion converges for any piecewise-continuous square-integrable function on the ball of radius $r_c$. The functions
\begin{equation*} 
\psi_{nlm}(r,\theta,\phi) = N^{(l)}_n j_l\bigg(r\frac{u_{ln}}{r_c}\bigg)Y^m_{l}(\theta,\phi)
\end{equation*}
where $N_n^{(l)}$ is a normalizing constant are known to form a complete orthonormal basis for square-integrable functions on this domain \cite{wang2009rotational,arvacheh2005pattern}. Since the proposed radial basis functions $g_{nl}(r)$ are derived by projecting the $j_l(r\frac{u_{ln}}{r_c})$ onto the space of functions with vanishing first and second derivatives at the boundary and constructing an orthonormal basis from the result, the basis functions in Eq.\ \ref{eq:expansion} constitute a complete orthonormal basis for square-integrable functions with the given boundary conditions as well.

With regard to the descriptors, completeness is usually considered to indicate whether the descriptors can be used to faithfully reconstruct a given local atomic environment up to symmetry. This paper instead defines completeness by whether the space of physically-distinct atomic environments is smoothly embedded by a map into the space of descriptors. This necessarily implies that there is an inverse map that allows the atomic environment to be reconstructed up to symmetry, and moreover that the map and its inverse are both continuous and differentiable. Observe that a local atomic environment with $\nu$ neighboring atoms is specified by $3\nu$ distinct relative spherical coordinates, but only $3(\nu - 1)$ quantities (e.g., the $\nu$ radial coordinates and $2\nu - 3$ triplet angles) are required to specify the environment up to rotations. This means that the space of physically-distinct atomic environments is $3(\nu -1)$-dimensional, and a complete set of descriptors maps this space to a $3(\nu -1)$-dimensional submanifold in the space of descriptors. Additionally, if the embedding is achieved using only the first $3(\nu -1)$ of the descriptors for any $\nu > 2$, then the descriptors are said to be optimally complete. Intuitively, an optimally complete set of descriptors encodes all relevant information (and just this information) about the atomic environment as concisely as possible.

There is limited discussion of completeness in the literature. One exception is the proof by Shapeev \cite{shapeev2016moment} that any rotation- and permutation-invariant polynomial can be written as a linear combination of the moment tensor descriptors, implying that these descriptors are complete (though probably not optimally complete). A recent dimensionality-reduction study \cite{imbalzano2018automatic} also touches on this question, attempting to optimize an MLP by reducing the dimension of the feature space. Possibility of such a reduction indicates that the BP and SOAP descriptors considered by the study contain substantial redundant information.

While proving that a set of descriptors is complete using the definition above is quite difficult, there is a necessary (but not necessarily sufficient) condition for completeness and optimal completeness that can be readily evaulated for any set of descriptors. This involves using the rank theorem \cite{krantz2012implicit} (a generalization of the implicit function theorem) to establish that the map from the space of physically-distinct atomic environments into the space of descriptors is locally invertible. More precisely, the condition establishes that for any particular atomic environment there is a set of closely-related and physically-distinct atomic environments over which the map into the space of descriptors is invertible, and that the inverse map is continuously differentiable. Practically speaking, this involves finding the rank $r$ of the Jacobian matrix $J$ of the function that transforms the relative atomic coordinates into the vector of descriptors. Assume for the moment that $r$ is constant. If $r < 3(\nu - 1)$, then the descriptors discard relevant information and cannot be complete. If $r > 3(\nu - 1)$, then the numerical calculation is faulty and should be checked. If $r = 3(\nu - 1)$, then the descriptors satisfy the necessary condition to be complete (though this local property does not necessarily extend to a global one). With regard to optimal completeness, let $J^{[q]}$ be the matrix formed by taking the first $q$ rows of $J$. If the rank of $J^{[3(\nu - 1)]}$ is $3(\nu - 1)$ for any $n > 2$, then the descriptors satisfy the necessary condition to be optimally complete.

Consider the $p_{nl}$ for the local atomic environment around the $i$th atom. From Eq.\ \ref{eq:pnl_reduced}, the $p_{nl}$ can be written as a function of the relative spherical coordinates of the neighboring atoms as
\begin{equation} 
p_{nl} = \frac{2l+1}{4\pi} \sum_{j} \sum_{k} g_{n-l,l}(r_{j}) g_{n-l,l}(r_{k}) P_l(\cos \theta_{j} \cos \theta_{k} + \sin \theta_{j} \sin \theta_{k} \cos (\phi_{j} - \phi_{k}))
\end{equation}
using various trigonometric identities. This defines a map from the 3$\nu$-dimensional space of relative spherical coordinates into the infinite-dimensional space of the descriptors $p_{nl}$. Let the Jacobian matrix of this map be constructed with rows labelled by the pairs $(n,l)$ in lexicographic order, where the $(n,l)$th row contains the $3 \nu$ partial derivatives of $p_{nl}$ with respect to the relative spherical coordinates (provided in the supplementary material). In practice, $J^{[q]}$ is constructed to consider the information content of only the first $q$ descriptors, and the rank of $J^{[q]}$ is found by performing singular value decomposition and counting the singular values that are substantially larger than the machine precision.

\begin{figure}
\center
\includegraphics[width=0.48\textwidth]{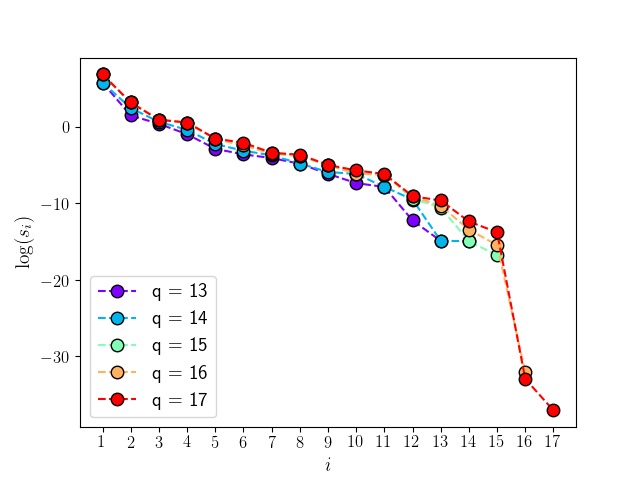}
\caption{The logarithmic singular values $\log(s_i)$ of $J^{[q]}$ for the $p_{nl}$ descriptors as a function of $q$. The decay of the $s_i$ to the machine precision for $i>15$ indicates that the rank of $J$ is 15.}
\label{fig:optimally_complete}
\end{figure}

\begin{figure}
\center
\includegraphics[width=0.48\textwidth]{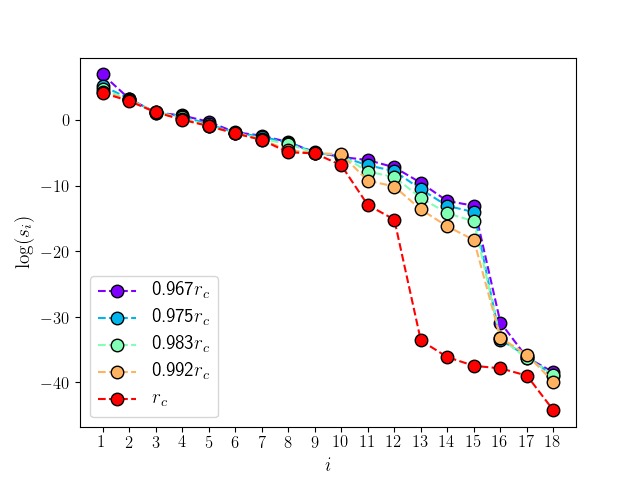}
\caption{The logarithmic singular values $\log(s_i)$ of $J^{[18]}$ for the $p_{nl}$ descriptors as a function of $q$. Five configurations were generated by scaling the initial configuration such that the radial coordinate of the most distant atom ranged from $0.967 r_c$ to $r_c$.}
\label{fig:atom_leaving}
\end{figure}

As a specific example, we generated an atomic environment with six randomly-positioned neighbors around a central atom, and constructed the $J^{[q]}$ for $13 \leq q \leq 17$. The significant and insignificant singular values $s_i$ are distinguished by plotting $\log(s_i)$ in Fig.\ \ref{fig:optimally_complete} and observing where the decay to machine precision occurs. The sharp drop after $q = 15$ clearly indicates that the rank of $J$ is $3(\nu - 1) = 15$, and similar results are obtained for different environments and different numbers of neighbors. This strongly suggests that the $p_{nl}$ satisfy the necessary conditions developed above for completeness and optimal completeness.

\begin{figure}
\center
\subfloat[]{%
	\label{subfig:bispectrum_standard}{%
		\includegraphics[width=0.48\textwidth]{%
			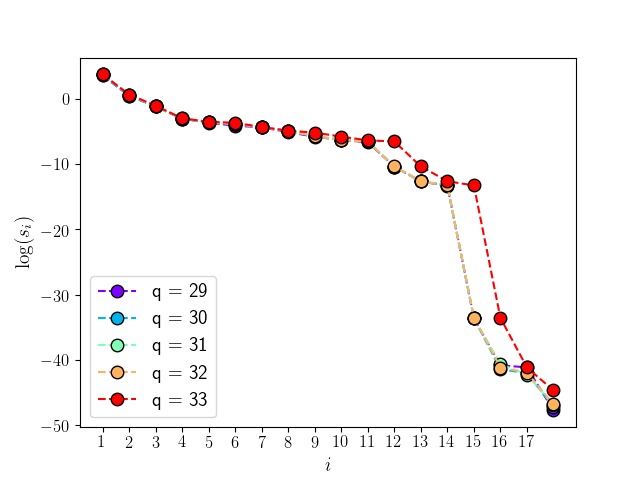}}} \quad
\subfloat[]{%
	\label{subfig:bispectrum_constrained}{%
		\includegraphics[width=0.48\textwidth]{%
			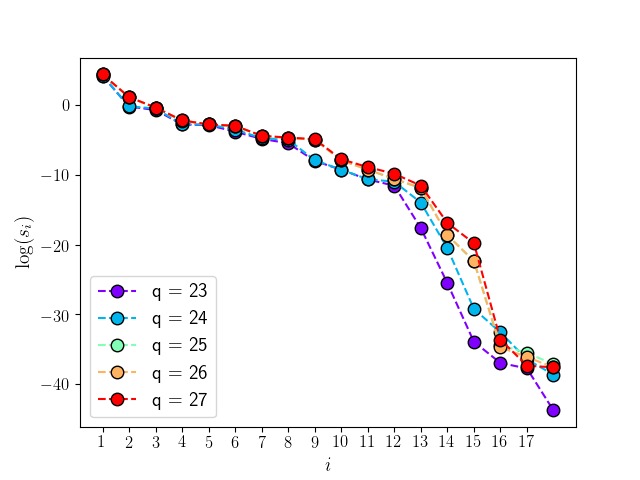}}}
\caption{\label{fig:bispectrum}The logarithmic singular values $\log(s_i)$ of $J^{[q]}$  for the $p_{n' n l}$ descriptors as a function of $q$. The figures correspond to descriptors indexed by (a) $0 \leq n' \leq n_{max}$, $0 \leq n \leq n_{max}$ and $0 \leq l \leq \min(n', n)$, and (b) the same except for $0 \leq n \leq n'$. The decay of the $s_{15}$ to the machine precision occurs much after $q > 15$, and this indicates that there are linearly dependent terms in both cases.}
\end{figure}

The above analysis is somewhat complicated by the differentiability of the $p_{nl}$ as atoms pass through the boundary at $r = r_c$; the differentiability of the $p_{nl}$ implies that the $s_i$ are continuous, and the number of significant $s_i$ should be reduced by three as an atom leaves the environment. This situation is considered in Fig.\ \ref{fig:bispectrum}, where $q$ is fixed at 18 and the $\log (s_i)$ are plotted as the most distant atom approaches the boundary. This shows that the rank behaves as expected, and moreover that the descriptors contain significant information even about atoms very close to the boundary.

While the calculation of the Jacobian matrices and evaulation of the completeness criterion for the other descriptors in the literature is outside the scope of this paper, we do consider the descriptors
\begin{equation}
p_{n' n l}=\sum_{m = -l}^{l} c_{n' l m}c^*_{n l m}
\label{eq:pnnl}
\end{equation}
where $0 \leq n' \leq n_{max}$, $0 \leq n \leq n_{max}$, and $0 \leq l \leq \min(n', n)$. This is motivated by the desire to explicitly couple information from radial basis functions with different indices $n'$ and $n$, and is closely related to the standard construction of the SOAP descriptors \cite{bartok2013representing,szlachta2014accuracy}. The Jacobian matrix of this map is constructed with rows labelled by the triplets $(n', n, l)$ in lexicographic order (partial derivatives are provided in the supplementary material). Figure \ref{subfig:bispectrum_standard} shows the logarithmic singular values of $J^{[q]}$ for an environment with six randomly-positioned neighbors. While these descriptors satisfy the same completeness condition as the $p_{nl}$ ($p_{nnl} = p_{nl}$) they are certainly not optimally complete, with $r = 15$ only for $q \geq 33$. The usual practice when constructing the SOAP descriptors \cite{imbalzano2018automatic} is to not place further restrictions on $n'$ and $n$, but inspection of Eq.\ \ref{eq:pnnl} indicates that $p_{n' n l} = p_{n n' l}$. That is, nearly half of the $p_{n' n l}$ and SOAP descriptors are trivially redundant and should be removed by enforcing the constraint $n \leq n'$. This is done in Fig.\ \ref{subfig:bispectrum_constrained}, and while the situation is improved ($r = 15$ for $q \geq 26$) there is still a significant number of redundant terms. Our conclusion is that constructing descriptors from the power spectrum with $n' = n$ is highly preferable to the $n' \neq n$ case.

\section{Performance and Efficiency}
\label{sec:efficiency}

\begin{figure}
\center
\includegraphics[width=10cm]{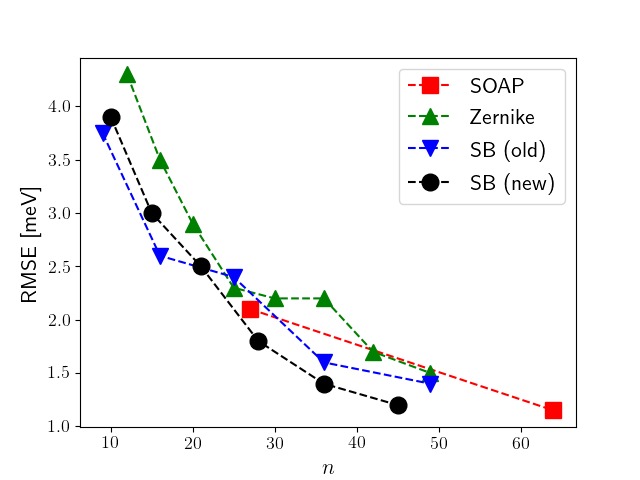}
\caption{The RMSE values on 1500 test configurations after 20000 training cycles obtained with ($n$-10-1) neural network architectures, where $n$ is the number of descriptors. The Stillinger-Weber analytical potential \cite{stillinger1985computer} was used in MD simulations of solid-state silicon at 1500 K to calculate the target energies.}
\label{fig:nnp}
\end{figure}

Since the intention for the SB descriptors is to be used as inputs for an MLP, a high-dimensional neural network potential (NNP) was constructed for solid-state silicon using the procedure described in Sections IIB and IIC of Kocer et al.\ \cite{kocer2019novel} The training data consisted of $8500$ environments sampled from an MD simulation equilibrated at $0$ Pa and $1500$ K that used the Stillinger--Weber potential \cite{stillinger1985computer}, and the error was defined as the root mean square deviation of the predicted energies of an additional $1500$ environments. Corresponding NNPs were constructed using the same data and training procedure for the prior SB descriptors \cite{kocer2019novel}, the SOAP descriptors \cite{szlachta2014accuracy}, and the Zernike descriptors \cite{khorshidi2016amp} (the BP descriptors having been considered previously \cite{kocer2019novel}), with the results reported in Fig.\ \ref{fig:nnp}. The slightly superior performance of the SB descriptors compared to the prior version is attributed to the elimination of the discontinuity at the origin of the basis functions and to the change in the order of summation in Eq.\ \ref{eq:expansion}, though this could be within the margin of error from the stochastic training of the neural network. The lower accuracy of the Zernike descriptors is attributed to the Zernike polynomials having the highest sensitivity to atomic positions in regions furthest from the central atom. The reason why there are different number of descriptors and data points of each type is that they all use different indexing schemes.

One of the advantages of the SB descriptors is that while they allow for the construction of MLPs of comparable accuracy to those using the SOAP descriptors, the SB descriptors are much faster to evaluate. Apart from the MATLAB implementations used for the timing experiments reported in Sec.\ \ref{sec:SB_desc}, a highly optimized C library with MATLAB and Python interfaces has been developed to calculate the SB descriptors and is publicly available\footnote{https://github.com/harharkh/sb\_desc}.

\section{Conclusion}

The descriptors used to describe a local atomic environment are one of the essential components of a machine learning potential. This paper identified a discontinuity in the basis functions used to construct the prior SB descriptors \cite{kocer2019novel} and the SOAP descriptors \cite{bartok2013representing}, and updated the indexing of the SB descriptors to make them twice-differentiable everywhere. Moreover, the SB descriptors were shown to satisfy a necessary condition for optimal completeness on the basis of the rank theorem \cite{krantz2012implicit}, establishing their ability to encode all relevant physical information about a local atomic environment using the fewest possible descriptors. At present, the SB descriptors are the only descriptors known to satisfy this condition. Moreover, they have been shown to be more than an order of magnitude faster to calculate than the SOAP descriptors, and an optimized code to calculate the SB descriptors has been made available. Finally, the performance of an NNP for solid-state silicon using the SB descriptors was compared to that of NNPs using the prior SB descriptors, the SOAP descriptors, and the Zernike descriptors.

\section{Supplementary Material}

See the supplementary material for a detailed derivation of the SB descriptors, of their derivatives with respect to the relative spherical coordinates of the surrounding atoms, and for MATLAB implementations of the SB descriptors and SOAP descriptors.

\section{Acknowledgments}

J.K.M. was supported by the National Science Foundation under Grant No. 1839370.


\bibliographystyle{unsrtnat}
\bibliography{refs.bib}

\end{document}